# A PARALLEL SELF–CONSISTENT FIELD CODE

LARS HERNQUIST[†], STEINN SIGURDSSON

Lick Observatory

Santa Cruz, California 95064

and

GREG L. BRYAN

National Center for Supercomputing Applications

5600 Beckman Institute

Department of Astronomy

University of Illinois at Urbana-Champaign

405 North Mathews Avenue

Urbana, Illinois 61801

## ABSTRACT

We describe a version of an algorithm for evolving self-gravitating collections of particles that should be nearly ideal for parallel architectures. Our method is derived from the "self-consistent field" (SCF) approach suggested previously by Clutton-Brock and others. Owing to the use of a global description of the gravitational field, the particles in an SCF simulation do not interact with one another directly, minimizing communications overhead between nodes in a parallel implementation. Ideal load-balancing is achieved since precisely the same number of operations are needed to compute the acceleration for each particle. Consequently, the SCF technique is perfectly scalable and the size of feasible applications will grow in simple proportion to advances in computational hardware.

We describe an SCF code developed for and tested on a Connection Machine 5. Empirical tests demonstrate the efficient and scalable nature of the algorithm. Depending on the application, simulations with particle numbers in the range $N \sim 10^7 - 10^{8.5}$ are now possible. Larger platforms should make simulations with billions of particles feasible in the near future. Specific astrophysical applications are discussed in the context of collisionless dynamics.

---

[†] Alfred P. Sloan Fellow, Presidential Faculty Fellow

The dynamics of many astronomical systems are well-approximated by the collisionless limit, in which encounters between, *e.g.*, individual stars are negligible. In particular, the two-body relaxation time of large galaxies is much longer than the age of the Universe (*e.g.* Binney & Tremaine 1987). Consequently, individual stellar orbits preserve certain integrals of motion to high precision, inhibiting the diffusion of orbits from one type to another.

Generally, the dynamics of galaxies are most easily studied using Monte Carlo simulations which represent phase space with $N$ discrete particles. As it is not yet possible to evolve $N \sim 10^{11} - 10^{12}$ particles numerically, computer models employ far fewer particles than the number of stars in a large galaxy. Thus, the relaxation time in computer simulations of galaxies is much shorter than in reality. To date, the same has also been true of computer models of dwarf galaxies which usually contain $\sim 10^6 - 10^7$ stars, although the algorithm described here makes calculations with particle numbers in this range practical. In any event, two-body effects, which are physically relevant to the evolution of small star clusters (*e.g.*, Spitzer 1987), are spurious in systems that are essentially collisionless.

The consequences of this enhanced relaxation can be subtle and are not easily detected by examining only the global properties of an object. For example, the integrity of individual orbits will no longer be preserved. In a perfectly smooth, stationary potential, the energy of any star will be exactly conserved. In a discrete system, however, the energy of any stellar orbit will random walk with a time-averaged diffusion rate $\propto N^{-1/2}$, even though the total energy of the system may be conserved to high precision. The other integrals of motion will behave similarly, depending on the approximations made in computing the forces on each particle. When $N$ is small compared to the actual number of stars in a galaxy, this "numerical diffusion" can thoroughly corrupt the dynamics of interest, although it may not lead to noticeable variations in the diagnostics typically used to gauge the accuracy of simulations, such as the total energy and angular momentum of the system.

For example, in a triaxial potential, a central mass concentration will scatter box orbits most effectively, as stars on these orbits eventually pass arbitrarily close to the origin (Gerhard & Binney 1985). If numerical relaxation allows stars to diffuse across orbital families, the extent to which a central mass will affect the structure of a triaxial system can be greatly exaggerated and the rate of, *e.g.*, feeding a central black hole will be unrealistically high (for a discussion see Binney & Petit 1989). Similarly, orbital diffusion can enhance the rate of tidal stripping in an external field (see, *e.g.*, Johnston & Hernquist 1994), rendering estimates of tidal disruption timescales meaningless. Even more severe are problems with "cold" dynamical systems such as disk galaxies, where relaxation can lead to a slow, monotonic conversion of ordered energy into random motions.

These types of processes occur on timescales which are short compared with conventional estimates of the two-body relaxation time since they do not require that the trajectories of all individual orbits suffer large deflections. Presently, there are no schemes for removing potential fluctuations by smoothing, owing to the global response of a system to discreteness noise (Weinberg 1993); the only remedy known is to employ a sufficiently large $N$ that individual orbits are integrated with the accuracy needed to preserve the quality of the dynamics under investigation. Algorithms such as the particle-mesh (PM) technique and tree-codes have made possible simulations of galaxies using $N \sim 10^5 - 10^6$, which is sufficient for certain problems. However, there are many applications where even this relatively large particle number is inadequate, particularly those phenomena which involve resonances (see, *e.g.*, Hernquist & Weinberg 1989, 1992).

Recently, there has been renewed interest in an approach termed the self-consistent field (SCF) method by Hernquist & Ostriker (1992). The basic idea underlying all SCF codes is to represent the potential of a self-gravitating system with a small number of basis functions, chosen to be optimal for, *e.g.*, investigating the global response of a stellar system to an external or internal perturbation. While this global decomposition of the potential does not greatly mitigate the effective rate of two-body relaxation relative to other N-body algorithms (Hernquist & Barnes 1989; Hernquist & Ostriker 1992; Weinberg 1993), the SCF technique is well-suited for parallel architectures. As we describe below, the intrinsic parallelizability of SCF codes together with the explosive growth of parallel computing now make practical simulations with particle numbers approaching the number of stars in small galaxies, so that relaxation is included in a nearly physical manner. This is not yet true of giant galaxies, although we expect that residual two-body effects will be negligible over many timescales of interest.

In this paper, we describe our adaptation of the SCF method for parallel computation. We have implemented this algorithm on a Connection Machine 5 (CM-5) and verified its inherently parallel nature. Our empirical tests show that practical simulations with $N \sim 10^7 - 10^{8.5}$ can now be performed. Since the SCF technique is also perfectly scalable, even larger simulations will be feasible in the near future. We anticipate that calculations of this size will lead to a qualitative change in our understanding of various aspects of the dynamics of collisionless systems.

The basic philosophy of the SCF technique is to solve Poisson's equation by expanding the density and potential in a set of basis functions. The expansion coefficients for the density are found by inversion, using the known density field carried by the particles. The expansion coefficients for the potential are obtained by solving Poisson's equation for the various basis functions. The acceleration for each particle is then calculated by analytically differentiating the potential expansion.

Slight variations in SCF codes result from the exact choice of basis functions and the technique used to solve for the coefficients in the potential expansion. For present purposes, we exclude those implementations in which some portion of Poisson's equation is solved directly (*e.g.* Hénon 1964; Aarseth 1967; van Albada & van Gorkom 1977; Fry & Peebles 1980; Villumsen 1982; White 1983; McGlynn 1984), and concentrate instead on schemes in which all coordinates are expanded in basis functions, since it is that approach which appears to be most promising for parallel calculations. By analogy with similar algorithms used to construct models of rotating stars (Ostriker & Mark 1968; Ostriker & Bodenheimer 1968), Hernquist & Ostriker (1992) chose to refer to this latter strategy as the self-consistent field method (a.k.a. "pure" expansions [Sellwood 1987]). The choice of basis functions is not unique, and depends on the coordinate system. Clearly, however, it is desirable to select a basis set so that the lowest order members provide a good approximation to the system being modeled so that it is not necessary to carry out the expansions to high order, compromising efficiency. A variety of basis sets have been tried (Clutton–Brock 1972, 1973; Aoki & Iye 1978; Allen *et al.* 1990; Hernquist & Ostriker 1992), although questions of efficiency and accuracy remain (see, *e.g.* Merritt & Trombley 1994).

In the discussion below, we are not concerned with the advantages or disadvantages of a specific basis set, but are interested solely in describing the use of SCF codes on parallel architectures. It is clear, *a priori*, that the SCF method will be nearly ideal for such hardware. Since the particles in these codes do not interact directly with one another, but only through their contribution to the global mean-field of the system, it is straightforward to show that the performance scaling is linear in the particle number. Effectively, then, the SCF method decomposes an $N$-body problem into $N$ one-body problems, which are trivial to parallelize.

*2.1 The approach*

For simplicity, consider a *biorthogonal* expansion, meaning that the basis functions for the density are orthogonal to those for the potential. (See, *e.g.* Saha 1993 for a more general formulation of the problem.)

In that event, we can write

$$\rho(\mathbf{r}) = \sum_{nlm} A_{nlm} \rho_{nlm}(\mathbf{r}), \tag{2.1}$$

and

$$\Phi(\mathbf{r}) = \sum_{nlm} A_{nlm} \Phi_{nlm}(\mathbf{r}), \tag{2.2}$$

where $n$ is analogous to a radial "quantum" number and $l$ and $m$ are analogous to angular "quantum" numbers. The use of biorthogonal basis sets yields a one–to–one relationship between the expansion terms for the density and potential and furthermore implies that the individual harmonics $\rho_{nlm}$ and $\Phi_{nlm}$ satisfy Poisson's equation

$$\nabla^2 \Phi_{nlm}(\mathbf{r}) = 4\pi \rho_{nlm}(\mathbf{r}). \tag{2.3}$$

In practice, the sums in equations (2.1) and (2.2) will be truncated at low orders, $n_{max}$ and $l_{max}$, and, as always, $m$ ranges between $-l$ and $+l$.

In what follows, we choose to illustrate the SCF method using the basis set derived by Hernquist & Ostriker (1992) for spheroidal galaxies. Empirical tests have shown that this algorithm can accurately reproduce solutions obtained by Vlasov solvers for spherical collapse (Hozumi & Hernquist 1994) and analytic results for the adiabatic growth of black holes in galaxies (Sigurdsson *et al.* 1994). Other basis sets would be more useful for, *e.g.* flat systems or objects whose density profiles are not well-approximated by de Vaucouleurs profiles. (For recent discussions of other basis sets see, *e.g.* Saha 1993; Qian 1992, 1993; Earn 1995; Earn & Sellwood 1995.)

For the specific case of a spheroidal mass distribution, it is natural to expand in spherical coordinates and use spherical harmonics to represent the angular dependence of the density and potential, so that equations (2.1) and (2.2) become

$$\rho(r, \theta, \phi) = \sum_{nlm} A_{nlm}\, \rho_{nl}(r)\, Y_{lm}(\theta, \phi) \tag{2.4}$$

and

$$\Phi(r, \theta, \phi) = \sum_{nlm} A_{nlm}\, \Phi_{nl}(r)\, Y_{lm}(\theta, \phi), \tag{2.5}$$

where, in general, the radial basis functions $\rho_{nl}(r)$ and $\Phi_{nl}(r)$ will depend on both the radial and polar quantum numbers. The spherical harmonics $Y_{lm}$ are defined in the usual manner (see, *e.g.*, Jackson 1975).

In their derivation, Hernquist & Ostriker (1992) took the zeroth order terms of expansions in equations (2.4) and (2.5) to be the spherical potential discussed by Hernquist (1990), namely

$$\rho_{000} \equiv \frac{1}{2\pi} \frac{1}{r} \frac{1}{(1+r)^3}, \tag{2.6}$$

and
$$\Phi_{000} \equiv -\frac{1}{1+r}, \tag{2.7}$$

where, for simplicity, the density and potential are expressed in dimensionless units.

Higher order terms are found by construction. First, it is assumed that terms with $n = 0$ but non-zero $l$ and $m$ behave asymptotically as would the corresponding terms in a multipole expansion of the potential. This yields

$$\rho_{0lm} = \frac{1}{2\pi} \frac{(2l+1)(l+1)}{r} \frac{r^l}{(1+r)^{2l+3}} \sqrt{4\pi} Y_{lm}(\theta, \phi) \tag{2.8}$$

and

$$\Phi_{0lm} \equiv -\frac{r^l}{(1+r)^{2l+1}} \sqrt{4\pi} Y_{lm}(\theta, \phi), \tag{2.9}$$

where the factor $\sqrt{4\pi}$ is included to give the correct normalization for $l = m = 0$.

The general terms with $n \neq 0$ are found by writing

$$\rho_{nlm}(\mathbf{r}) \equiv \frac{K_{nl}}{2\pi} \frac{r^l}{r(1+r)^{2l+3}} W_{nl}(\xi) \sqrt{4\pi} Y_{lm}(\theta, \phi), \tag{2.10}$$

and

$$\Phi_{nlm}(\mathbf{r}) \equiv -\frac{r^l}{(1+r)^{2l+1}} W_{nl}(\xi) \sqrt{4\pi} Y_{lm}(\theta, \phi), \tag{2.11}$$

where $K_{nl}$ is a constant and the functions $W_{nl}(\xi)$ include the remaining ($n > 0$) radial dependence and are found by inserting equations (2.10) and (2.11) into Poisson's equation (2.3). The resulting expression is simplified by the transformation of variables

$$\xi = \frac{r-1}{r+1}, \tag{2.12}$$

which reduces the radial equation to a Sturm-Liouville form with well-known solutions known as ultraspherical or Gegenbauer polynomials, $C_n^{(\alpha)}(\xi)$ (e.g. Szegö 1939; Sommerfeld 1964). Inserting equations (2.10) and (2.11) into Poisson's equation, some algebra (Hernquist & Ostriker 1992) enables us to infer

$$W_{nl}(\xi) \equiv C_n^{(2l+3/2)}(\xi), \tag{2.13}$$

and

$$K_{nl} = \frac{1}{2} n(n + 4l + 3) + (l+1)(2l+1), \tag{2.14}$$

so that the full basis sets for the density and potential are

$$\rho_{nlm}(\mathbf{r}) = \frac{K_{nl}}{2\pi} \frac{r^l}{r(1+r)^{2l+3}} C_n^{(2l+3/2)}(\xi) \sqrt{4\pi} Y_{lm}(\theta, \phi), \tag{2.15}$$

and

$$\Phi_{nlm}(\mathbf{r}) = -\frac{r^l}{(1+r)^{2l+1}} C_n^{(2l+3/2)}(\xi) \sqrt{4\pi}\, Y_{lm}(\theta, \phi), \tag{2.16}$$

where the $K_{nl}$ are defined in equation (2.14).

It is straightforward to show that these expansions are indeed biorthogonal; *i.e.*

$$\int \rho_{nlm}(\mathbf{r})\, [\Phi_{n'l'm'}(\mathbf{r})]^*\, d\mathbf{r} \;=\; I_{nl}\, \delta_{l'l}\, \delta_{m'm}\, \delta_{n'n}, \tag{2.17}$$

where

$$I_{nl} = -K_{nl}\, \frac{4\pi}{2^{8l+6}}\, \frac{\Gamma(n+4l+3)}{n!\,(n+2l+3/2)\,[\Gamma(2l+3/2)]^2} \tag{2.18}$$

(Hernquist & Ostriker 1992). Thus, for a known density profile $\rho(\mathbf{r})$ the expansion coefficients $A_{nlm}$ appearing in equations (2.1) and (2.2) can easily be obtained by multiplying both sides of equation (2.1) by $[\Phi_{n'l'm'}(\mathbf{r})]^*$ and integrating. This gives

$$\int \rho(\mathbf{r})\, [\Phi_{n'l'm'}(\mathbf{r})]^*\, d\mathbf{r} = \sum_{nlm} A_{nlm} I_{nl}\, \delta_{l'l}\, \delta_{m'm}\, \delta_{n'n} = A_{n'l'm'} I_{n'l'}, \tag{2.19}$$

or, relabeling,

$$A_{nlm} = \frac{1}{I_{nl}} \int \rho(\mathbf{r})\, [\Phi_{nlm}(\mathbf{r})]^*\, d\mathbf{r}, \tag{2.20}$$

where the integration is over all space.

For application to $N$-body simulations the density field will be represented by $N$ discrete particles and equation (2.20) can be written

$$A_{nlm} = \frac{1}{I_{nl}} \sum_{k=1}^{N} m_k\, \Phi_{nl}^*(r_k)\, Y_{lm}^*(\theta_k, \phi_k), \tag{2.21}$$

where $m_k$ and $(r_k, \theta_k, \phi_k)$ are the mass and coordinates of the $k$-th particle. Once all the $A_{nlm}$ have been calculated from the known coordinates of all the particles, the potential at the location of any particle can be evaluated using (2.5)

$$\Phi(r_k, \theta_k, \phi_k) = \sum_{nlm} A_{nlm}\, \Phi_{nl}(r_k)\, Y_{lm}(\theta_k, \phi_k). \tag{2.22}$$

The components of the acceleration can then be obtained by simple analytical differentiation of equation (2.22).

*2.2 Intrinsic parallel structure*

The parallel nature of the various steps in the SCF method are best appreciated by writing the expressions in equations (2.21) and (2.22) symbolically as nested DO loops. Thus, the section of code to compute the expansion coefficients will look like

DO $n = 0, n_{max}$

    DO $l = 0, l_{max}$

        DO $m = -l, l$

            DO $k = 1, N$

                compute contribution of particle $k$ to $A_{nlm}$

            ENDDO

        ENDDO

    ENDDO

ENDDO

Similarly, the potential (and acceleration) for each particle will be computed in four nested loops:

DO $k = 1, N$

    DO $n = 0, n_{max}$

        DO $l = 0, l_{max}$

            DO $m = -l, l$

                compute contribution of basis function $nlm$ to

                potential and acceleration of particle $k$

            ENDDO

        ENDDO

    ENDDO

ENDDO

Nearly all ($\gtrsim 95\%$) the cpu time required by an SCF simulation is used to process structures like those sketched above. The four loops are completely interchangeable, and their optimal ordering depends on the hardware. On a vector machine, such as a Cray, it is desirable that the innermost loop be the longest one; in this case the one over $k$. While this choice results in considerable enhancement in performance on vector platforms, it also leads to a severe penalty in memory overhead. For example, with the loops arranged as in the first symbolic block of code above, it is most efficient to compute quantities depending on only $n$ and $l$ outside the loop over $m$. This can be achieved by using temporary arrays of length $N$ to store these quantities for later use in the loop over the particles. Using this approach, Hernquist & Ostriker (1992) obtained nearly an order of magnitude gain in speed on a Cray C-90, but their code needed 30 arrays of length $N$ to store all the data associated with the particles.

On serial or parallel platforms, however, this organization is wasteful. In fact, it is most expedient to order the loops so that the *outermost* one is over the particles, to minimize memory requirements. Thus, all the time-consuming sections of a parallel SCF code will be organized similarly to the second block of code above. By further combining the routines which evaluate the force and integrate particle coordinates, it is possible to eliminate *all* extraneous arrays. The memory usage of such an SCF code will be either $\approx 6N$ or $\approx 7N$ if a simple leapfrog integrator is employed. Clearly, six arrays are needed to store permanent copies of the phase space. One additional array of length $N$ is required if the particles have a spectrum of masses. In some situations, depending on the optimization or other physical variables required, it may also be necessary to use one or two additional temporary arrays of length $N$.

It is also clear from the symbolic code above that the SCF method scales with particle number as $\sim O(N)$, since the loops involve simple linear passes through the particles. This assumes, of course, that the number of expansion terms, $(n_{max} + 1)(l_{max} + 1)(2l_{max} + 1)$, is small compared with $N$.

As noted above, for parallel architectures the two sections of code requiring nearly all the cpu time in an SCF calculation are most naturally ordered as

DO $k = 1, N$

    DO $n = 0, n_{max}$

        DO $l = 0, l_{max}$

            DO $m = -l, l$

                compute contribution of particle $k$ to $A_{nlm}$  $or$

                calculate contribution of basis function $nlm$

                to potential and acceleration of particle $k$

            ENDDO

        ENDDO

    ENDDO

ENDDO

The highly parallel nature of the SCF algorithm is readily apparent from this structure. Suppose that the platform has $p$ nodes so that $p << N$ (more generally $p = \#$ processors $\times \#$ computing units per processor [*e.g.* the vector units on the CM5] $\times$ length of the vector pipeline of each computing unit). The basic strategy then is to assign $N/p$ particles to each node. The first stage of an SCF force evaluation is parallelized by accumulating the partial contribution to each $A_{nlm}$ from the $N/p$ particles associated with each node, in parallel over all the nodes. The step is perfectly parallelizable, requires no communications among processors, and, for suitable choices of $N$ and $p$, provides ideal load-balancing. Once the partial contributions have been calculated, the full $A_{nlm}$ are found by summing over all processors. The $A_{nlm}$, gathered on a host machine or a nominal master processor, are then broadcast to each node. These steps require communications between processors, but are negligible in cost for low communications overhead since the number of nodes is assumed to be small compared with $N$. Once copies of the $A_{nlm}$ have been distributed to all the processors, the potential and force on each particle is obtained by summing over the basis functions. This step, which is time-consuming, again requires no interprocessor communication and provides ideal load-balancing. Finally, the phase-space coordinates of each particle are updated on the processor to which it is assigned using the locally calculated forces.

The various steps in an ideal parallel SCF simulation can thus be concisely summarized as follows:

1) Initially, map $N$ particles onto $p$ nodes, $N/p$ particles per node. Thereafter, each particle is permanently associated with the node to which it is assigned. Depending upon how the initial conditions are provided, this step may require communication between the nodes and the host computer, but it is done only once per simulation.

2) Initialize the local variables associated with each node, including copies of the $A_{nlm}$. In general, this step will also require some communication between the host and the nodes, but its cost will be negligible.

3) In parallel over all the nodes, compute the partial contributions to the $A_{nlm}$ from the $N/p$ particles assigned to each processor. This step, which is one of two which dominate the cpu costs, is perfectly parallelizable, in principle requires no communication, and provides ideal load balancing if $N/p$ is an integer.

4) Sum over all nodes to compute the full $A_{nlm}$, temporarily storing the results on the host machine. This requires interprocessor communication as well as communication between the host and the nodes, but can be done in a time $\propto \log p$ and will be completely negligible in the practical case where $p << N$ and latency is not very large.

5) Broadcast the full values of the $A_{nlm}$ from the host back to all the nodes. Again, this step requires communications but will be negligible in cost for the reasons given in step 4.

6) In parallel over all the nodes, compute the potential and acceleration for each particle by summing over the relevant expansions. This is the second step which dominates the cpu costs. Like step 3, it is perfectly parallelizable, in principle requires no communication, and provides ideal load balancing if $N/p$ is an integer. Note that the force and potential for a given particle are computed locally on the node to which it was initially assigned.

7) In parallel over all the nodes, advance the phase-space coordinates using the acceleration computed for each particle. Unless the number of expansion terms is very small, this step will be negligible in cost. For the same reasons noted in steps 3 and 6, this step is also perfectly parallelizable, requires no communications, and achieves ideal load balancing if $N/p$ is an integer.

8) Occasionally, but not often, output data to disk. Clearly, this step can be quite slow, and it will be necessary to consider schemes for optimizing data storage if $N$ is large. As discussed below, a good strategy may be to save the $A_{nlm}$ frequently, but only write-out subsets of the phase-space data, doing most of the analysis as part of the actual simulation.

9) Cycle back to step 3, and iterate for as many timesteps as is necessary to complete the simulation.

Provided that care is taken in dealing with the transfer of data from mass disk to the processors, the only steps above which will require significant cpu time are steps 3 and 6. As these involve nothing more than evaluating the four nested DO loops indicated at the beginning of §3, it is apparent that the SCF method is nearly optimal from the point of view of parallel computing. Aside from steps which require negligible cost, the SCF method is ideally parallelizable in principle in that it requires virtually no communications, achieves exact load balancing, and is perfectly scalable. In practice, some overhead can be introduced by compiler inefficiency into steps 3 and 6, as discussed at the end of §4 below.

The only aspect of a calculation as outlined symbolically in steps 1 through 9 above that might present a bottleneck on the parallel nature of the algorithm is step 8, when data is written to disk. Fast data transfer is essential for an efficient parallel implementation of the algorithm. Simulations with $\gtrsim 10^7$ particles need $\gtrsim O(10^9)$ bytes of storage for each snapshot of phase space stored. The NCSA CM-5 has a 135 Gb scalable disk array (SDA), with peak transfer rate of $\sim$ 120 Mb/sec and sustained transfer rate of $\sim$ 90 Mb/sec. The SDA provides disk–striped, big–endian IEEE floating point format. Files are accessed through special function calls, providing either serial–order or random access. In practice, loading a 512 Mb serial–order file into 256 nodes requires approximately 1.5 CPU seconds and less than 7 seconds total, writing the file is a little slower (see also Kwan & Reed 1994). This is fast enough for input and output not to be a significant bottleneck, provided output is no more frequent than every few hundred integration steps. A 512 Mb file stores a single snapshot of 8,388,608 particle phase space, and the Unix file size limit of 2.1 Gb will soon be a serious constraint on storing even single vectors once $N \gtrsim 10^{8.5}$. As we note in section 5, these considerations require a basic rethinking of how N–body simulations will be done as $N$ approaches $O(10^9)$.

We have developed a parallel SCF code, based on the principles outlined in §3. As an illustration, we have used the basis set proposed by Hernquist & Ostriker (1992), as described in §2.1. We emphasize that this choice is not unique, and is certainly not the most appropriate one in all cases, but a variety of empirical tests have demonstrated that it can reproduce well-known analytical and numerical solutions (*e.g.* Hozumi & Hernquist 1994; Quinlan *et al.* 1994; Sigurdsson *et al.* 1994). Moreover, the essential aspects of the parallelization of SCF codes described here should apply to other basis sets as well.

The details of our parallel SCF code are similar to those of the vector algorithm of Hernquist & Ostriker (1992). For example, special functions are again evaluated using recursion relations, although we no longer compute the functions for all the particles simultaneously since we do not need to satisfy constraints imposed by vector processors. The parallel code is simple and concise. For these reasons, it is being ported to a variety of platforms with relative ease. As a specific illustration, we describe tests done with the parallel SCF code on the 512 processor CM-5 at the National Center for Supercomputing Applications. This machine supports a total of 16 Gbytes of memory. Hence, as noted in §3, the ideal SCF algorithm running on this machine will be limited to simulations with particle numbers $N \lesssim 3 \times 10^8$ or $N \lesssim 6 \times 10^8$ in double or single precision mode, respectively. We have verified that the code produces correct results by comparing calculations done on the CM-5 with others done on the Cray C-90 at the Pittsburgh Supercomputing Center and serial machines at UCSC using the old implementation of Hernquist & Ostriker (1992). The results agree to the roundoff accuracy of the various platforms.

Table 1 shows timing tests done at the NCSA 512 node CM-5 and comparison tests done on the 32 node CM-5 at the University of Maryland and the 88 node Intel Paragon at Indiana University (Yang *et al.* 1994, Gannon *et al.* 1994). The original version of the algorithm, run on 512 nodes at NCSA CM-5 reached approximately 14.4 Gflops sustained, already making it one of the most efficient applications running on this machine (see Hillis & Boghosian 1993). The flop rate was calculated by comparing the basic algorithms running on a Cray–YMP at a fixed $n_{max}, l_{max}$. Since then, compiler improvement and minor code modifications have increased the performance by about 20%, and the current version of the code reaches a sustained peak of about 18 Gflops on 512 nodes with $N \gg 10^6$. The basic SCF algorithm requires 7850 floating points operations per step (for $n_{max} = 6$, $l_{max} = 4$). The timing tests presented in Table 1 are for a version of the code where an external potential is included, requiring some additional scalar overhead and $\sim 100$ extra floating point operations per step. The pC++ version of the code requires almost twice the memory, but is about 10% faster than the original code. For 51,200 particles, $n_{max} = 6$, $l_{max} = 4$, the

pC++ code takes 0.46 seconds per timestep on a 32 node CM-5, and 3.3 seconds per time step on a 32 node Intel Paragon (running non-optimized F77 subroutines) (Yang *et al.* 1994, Gannon *et al.* 1994).

The scaling of the code is slightly superlinear with increasing $N$ for $N/p$ small, as the scalar overhead per time step is only weakly dependent on $N$. The scaling with increasing $n_{max}$ is slightly worse than linear, while the scaling with increasing $l_{max}$ is slightly superlinear in $l_{max}^2$, somewhat better than expected. With 4 vector units per node, we require $O(100,000)+$ particles on a 64 node partition to see the linear scaling with $N$ expected. We do not present tests for larger $N$, as the generation of initial conditions and file handling was cumbersome, and required scheduling dedicated time. The installation of the HIPPI network at the NCSA, providing high speed parallel file transfer, will make the handling of larger simulations much easier.

We also show the performance of the multi-step scheme discussed in Sigurdsson *et al.* (1994). For a particle system with potential $\Phi = \Phi_{SCF} + \Phi_{ext}$, where $\Phi_{SCF}$ is the potential due to the self-gravity of the particles, and $\Phi_{ext}$ is some time variable external potential, we update $\Phi_{ext}(r_i, t)$ and the particle positions at time intervals, $dt_s \ll dt$, holding $\Phi_{SCF}$ fixed. This is a physically good approximation if the response of the particles to the external potential produces slow changes in the global distribution and rapid changes locally. In some cases the multi-step scheme can provide enormous speed-up of the integration compared with simply decreasing $dt$, as can be seen from column 6 in Table 1, shown here for $\Phi_{ext}$ as the Kepler potential due to a central massive black hole. Not until $dt/dt_s = 1024$, does the multi-step scheme approach diminishing return, where the effort spent updating the black hole potential becomes comparable to computing the particle potential. The scheme permits us to use smaller smoothing length for the black hole potential, increasing significantly the resolution of the code at small spatial scales.

The code as implemented achieves about 28% of the theoretical maximum flop rate on a 512 node CM-5. This is quite a respectable performance by the standards of parallel algorithms (Hillis & Boghosian 1993), especially considering that no low-level optimization has been done. Two factors preclude us from approaching theoretical sustained peak performance on the CM-5. The peak rate requires each vector unit execute a concurrent multiply-add per cycle. That is an algorithm must perform a multiply, followed by an add using the outcome of the multiple, each cycle. While portions of our code do execute optimally in this sense, some divides and subtracts are inevitable, and as structured the code cannot exceed about 1/3 peak. The restructured algorithm, where the loop over particles is outermost, reaches higher peak floating point rate, but on the CM-5 the host node calculates the memory offset for each load into cache, which leads to $\sim 50\%$ communications overhead. Unrolling the loops over $n, l$ and $m$ recovers the sustained flop rate, with somewhat shorter time per integration step needed. A pre-compiler that unrolls fixed loops would be useful.

The resulting flop rate has not been calculated and cannot be directly compared with the vector optimized code as some rearrangement of low level recursion relations has been made.

Stars in collisionless systems move along orbits essentially unperturbed by individual interactions with one another. Computer models of such objects will faithfully represent their dynamics only to the extent that two-body effects are negligible over the time-scales of interest. At present, high accuracy can only be achieved by employing large numbers of particles in the numerical simulations.

In this paper, we have described an implementation of the SCF method for parallel architectures. Empirical tests on a CM-5 verify the intrinsic parallel nature of this approach and demonstrate that simulations with particle numbers up to $N \sim 10^{8.5}$ are now possible, which greatly exceeds the size of any present serial or vector application in galactic dynamics. While still small compared to the actual number of stars in a giant galaxy, such values of $N$ are likely sufficient for investigating the detailed dynamics of many problems of current interest.

Among the problems of immediate interest are: the response of dwarf galaxies to tidal perturbations (*e.g.* Kuhn & Miller 1989; McGlynn 1990; Johnston & Hernquist 1994; Johnston, Spergel & Hernquist 1994), the growth of central black holes in galaxies (*e.g.* Norman *et al.* 1985; Hasan & Norman 1990; Sigurdsson *et al.* 1994), the dynamics of binary black holes in galaxies (*e.g.* Begelman *et al.* 1980; Ebisuzaki *et al.* 1991; Quinlan & Hernquist 1994), instabilities in galaxy models (*e.g.* Merritt & Aguilar 1985; Barnes *et al.* 1986; Palmer & Papaloizou 1988; Merritt & Hernquist 1991), the orbital decay of satellites (*e.g.* White 1983; Lin & Tremaine 1983; Bontekoe 1988; Hernquist & Weinberg 1989), coupling between bars and spheroids (*e.g.* Little & Carlberg 1991a,b; Hernquist & Weinberg 1992), and the dynamics of multiple black hole systems in stellar backgrounds (*e.g.* Kulkarni *et al.* 1993; Sigurdsson & Hernquist 1993). Previous applications in these areas have generally employed particle numbers in the range $N \sim 10^3 - 10^5$. With algorithms like those described here, and straightforward extensions of these algorithms, these problems can now be attacked with roughly a ten thousand-fold increase in $N$, which should lead to markedly improved results in cases where an accurate handling of individual orbits is essential, or when a complete coverage of phase space is crucial for densely populating resonances. In addition, larger particle numbers extend the dynamic range in masses and density probed by simulations and makes it possible to explore gradients of projected quantities over greater spatial scales, owing to the reduction of sampling noise.

Because of its simple structure, the parallel SCF algorithm should port easily to other platforms. Among the most promising are the Cray T3D and the Intel Paragon. A portable version of our algorithm written in pC++ has already been developed by Gannon *et al.* (1994) and is now being tested on a variety of platforms, including those just noted as well as heterogeneous systems (Melhem & He [private communication]). A ten-fold increase in $N$ over what is now possible on the CM-5 should be attainable in the immediate future, making possible simulations employing billions of particles.

Computations of this size require a rethinking of the basic approach to doing numerical simulations. To see this, consider the amount of data that would be generated by a double-precision calculation with $N = 10^9$. A binary file containing the essential particle information, their masses and phase space coordinates, would be $7 \times 8 \times 10^9 = 56$ Gbytes in length. A simulation generating dozens of these snapshots would produce in excess of 1 terabyte of data. The prospect of transferring, reducing, analyzing, and archiving data-sets of this size is indeed daunting.

Computer simulations consist of three stages: 1) generating initial conditions, 2) evolving the system with an evolution code, and 3) reducing and analyzing the output of the calculation. Up to now, it has proved most convenient to treat these phases independently. Given the amount of data that would be generated by large computations on massively parallel architectures, this simple division of tasks will no longer be practical. Evidently, the simulation code will be required to generate initial conditions, evolve the system, perform analysis, and reduce the data in a form that does not compromise the ability of the user to glean scientific insight from the computer model.

In fact, in our study of the adiabatic growth of black holes in galaxies, we have already taken steps in this direction (Sigurdsson *et al.* 1994). Our SCF code has the capability to generate $N$-body realizations of simple distribution functions in parallel. Analysis and data reduction are performed as the simulation proceeds and output may be restricted to subsets of the phase space information carried by the particles. In this latter function, we are aided by the concise representation of the system offered by the basis function expansions. It is practical to store the evolution of the $A_{nlm}$ densely in time. Thus, a relatively coarse sampling of the particle data can be output and used, *after the fact*, to perform an orbital analysis by evolving these particles in the known time-dependent potential field determined from the SCF expansions. This operation would be considerably more difficult, if not impossible, for applications using tree-codes or PM algorithms.

$N$-body simulations have progressed far in the more than 20 years that have elapsed since the study of bar instabilities in disks by Ostriker & Peebles (1973), in which individual galaxies were represented with only a few hundred particles. Current state-of-the-art computer models of galaxies employ tens or hundreds of thousands of self-gravitating particles. Nevertheless, considerable skepticism remains over the interpretation of many $N$-body simulations, owing to the elevated levels of two-body relaxation present in the models compared to real galaxies (for discussions see, *e.g.*, Sellwood 1987, Hernquist & Ostriker 1992, Weinberg 1993). Numerical diffusion of orbits driven by potential fluctuations can perturb the structure of equilibria, broaden resonances, artificially enhance evaporation processes, and isotropize velocity distributions on unrealistically short time-scales. Algorithms like the parallel SCF code described here offer the possibility of greatly suppressing these effects by enabling the use of a factor $\sim 10^4$ larger particle number over most previous applications. In addition to reducing discreteness noise by a factor $\sim 100$, such an increase in $N$ would permit convergence studies to be performed over a much larger baseline in particle number. Advances such as these will undoubtedly improve the rigor of $N$-body models of galaxies and make it practical to use these tools to probe collisionless dynamics in greater detail than has yet been possible.

ACKNOWLEDGEMENTS

This work was supported in part by the Pittsburgh Supercomputing Center, the National Center for Supercomputing Applications, the Alfred P. Sloan Foundation, NASA Theory Grant NAGW–2422, NSF Grants AST 90–18526 and ASC 93–18185, and the Presidential Faculty Fellows Program.

Table 1:

Execution time per timestep for different particle numbers, $N$, expansion co–efficients, $n_{max}$, $l_{max}$ and number of nodes. The time shown is for a single integration step, except for where the multi–step schemes indicated in the last column, for which the times shows are for a "big" time step. The last column shows the time per integration step per particle, $\times p/[(n_{max}+1)(l_{max}+1)(2l_{max}+1)]$, illustrating the scaling of the code for different $N, p$ and co–efficients. The code is slightly super–linear in $N$ for the $N/p$ used here, as the fractional scalar overhead is relatively smaller for the larger $N$.

Table 1.

| $N$ | p | $n_{max}$ | $l_{max}$ | $t$ per $dt$ secs | $dt/dt_s$ | $t_n$ $\mu$sec |
|---|---|---|---|---|---|---|
| 51,200 | 32 | 6 | 4 | 0.46 | – | 0.91 |
| 512,000 | 64 | 6 | 4 | 2.25 | – | 0.89 |
| | | 6 | 0 | 0.22 | – | 0.39 |
| | | 16 | 0 | 0.62 | – | 0.46 |
| | | 16 | 4 | 8.23 | – | 1.34 |
| | | 16 | 9 | 31.2 | – | 1.21 |
| | 128 | 6 | 4 | 1.17 | – | 0.93 |
| | 256 | 16 | 9 | 8.67 | – | 1.34 |
| 512,000 | 64 | 16 | 0 | 0.62 | 1 | 0.46 |
| | | | | 0.65 | 4 | 0.48 |
| | | | | 0.78 | 16 | 0.58 |
| | | | | 0.97 | 32 | 0.72 |
| | | | | 2.05 | 128 | 1.52 |
| | | | | 3.48 | 256 | 2.58 |
| | | | | 12.1 | 1024 | 8.98 |
| 8,388,608 | 256 | 6 | 4 | 9.30 | – | 0.90 |
| | 256 | 16 | 9 | 121 | – | 1.14 |